\documentstyle[]{l-school}

\def\a{\alpha}
\def\b{\beta}

\def\g{\gamma}

\def\L{\Lambda}

\def\del{\partial}

\def\beq{\begin{equation}}
\def\eeq{\end{equation}}
\def\bea{\begin{eqnarray}}
\def\eea{\end{eqnarray}}

\begin{document}

\markboth{David G. Robertson}{PHYSICAL COUPLING SCHEMES\dots}

\setcounter{part}{10}

\title{PHYSICAL COUPLING SCHEMES AND QCD EXCLUSIVE PROCESSES}

\author{David G. Robertson}

\institute{Department of Physics\\
The Ohio State University\\
Columbus, OH 43210, USA}

\maketitle

\noindent
One of the most important problems in making reliable predictions in
perturbative QCD is dealing with the dependence of the truncated
perturbative series on the choice of renormalization scale $\mu$ and
scheme $s$ for the QCD coupling $\alpha_s(\mu)$.  Consider a physical
quantity ${\cal O}$, computed in perturbation theory and truncated at
next-to-leading order (NLO) in $\alpha_s$:
\beq
{\cal O} = \a_s(\mu)\left[1+\left(A_1+B_1 n_f\right)
{\a_s(\mu)\over\pi} + \cdots\right]\; ,
\label{1}
\eeq
where $n_f$ is the effective number of quark flavors.  The
finite-order expression depends on both $\mu$ and the choice of scheme
used to define the coupling.  In fact, Eq. (\ref{1}) can be made to
take on essentially any value by varying $\mu$ and the renormalization
scheme, which are {\em a priori} completely arbitrary.  The
scale/scheme problem is that of choosing $\mu$ and the scheme $s$ in
an ``optimal'' way, so that an unambiguous theoretical prediction,
ideally including some plausible estimate of theoretical
uncertainties, can be made.\footnote{The precise meaning of
``optimal'' in this context is connected to the minimization of
remainders for the truncated series.  As is well known, perturbation
series in QCD are asymptotic, and thus there is an optimum number of
terms that should be computed for a given observable.  In general,
very little is known about the remainders in pQCD; however, if we
assume that pQCD series are sign-alternating, then the remainder can
be estimated by the first neglected (or last included) term.  This
term can take on essentially any value, however, by simply varying the
scale and scheme, and thus its minimization is meaningless without
invoking additional criteria.  }

For any given observable there is no rigorously correct way to make
this choice in general.  However, a particular prescription may be
supported to a greater or lesser degree by general theoretical
arguments and, {\em a posteriori}, by its success in practical
applications.  From these perspectives, a particularly successful
method for choosing the renormalization scale is that proposed by
Brodsky, Lepage and MacKenzie \cite{BLM}.  In the BLM procedure, the
renormalization scales are chosen such that all vacuum polarization
effects from fermion loops are absorbed into the running couplings.  A
principal motivation for this choice is that it reduces to the correct
prescription in the case of Abelian gauge theory.  Furthermore, the
BLM scales are physical in the sense that they typically reflect the
mean virtuality of the gluon propagators.  Another important advantage
of the method is that it ``pre-sums" the large and strongly divergent
terms in the pQCD series which grow as $n!  (\alpha_s \beta_0 )^n$,
i.e., the infrared renormalons associated with coupling constant
renormalization.

Dependence on the renormalization scheme can be avoided by considering
relations between physical observables only.  By the general
principles of renormalization theory, such a relation must be
independent of any theoretical conventions, in particular the choice
of scheme in the definition of $\alpha_s$.  A relation between
physical quantities in which the BLM method has been used to fix the
renormalization scales is known as a ``commensurate scale relation''
(CSR) \cite{CSR}.  An important example is the generalized Crewther
relation \cite{CSR,BrodskyKataevGabaladzeLu}, in which the radiative
corrections to the Bjorken sum rule for deep inelastic lepton-proton
scattering at a given momentum transfer $Q$ are predicted from
measurements of the $e^+ e^-$ annihilation cross section at a
commensurate energy scale $\sqrt s \propto Q$.

In this talk I summarize recent applications of the BLM procedure to
obtain CSRs relating QCD exclusive amplitudes to other observables, in
particular the heavy quark potential $V(Q^2)$ \cite{bjpr97}.  As we
shall see, the heavy quark potential can be used to define a physical
coupling scheme which is quite natural for perturbative calculations.
It may also be useful in the context of a nonperturbative formulation
of QCD based on light-cone quantization.

\section*{BLM SCALE FIXING}

The term involving $n_f$ in Eq. (\ref{1}) arises solely from quark
loops in vacuum polarization diagrams.  In QED these are the only
contributions responsible for the running of the coupling, and thus it
is natural to absorb them into the definition of the coupling.  The
BLM procedure is the analog of this approach in QCD.  Specifically, we
rewrite Eq. (\ref{1}) in the form
\beq
{\cal O} = \a_s(\mu)\left[1 + 
\left({3\b_0 B_1\over2}\right) {\a_s(\mu)\over\pi}
\right]
\left[1 + 
\left(A_1+ {33B_1\over2}\right){\a_s(\mu)\over\pi}
\right]\; ,
\eeq
correct to order $\a_s^2$, where $\b_0 = 11-{2n_f/3}$ is the
lowest-order QCD beta function.  The first term in square brackets can
then be absorbed by a redefinition of the renormalization scale in the
leading-order coupling, using
\beq
\a_s(\mu^*) = \a_s(\mu)\left[1-{\b_0\a_s(\mu)\over2\pi}
\ln\left(\mu^*/\mu\right)+ \cdots\right]\; .
\eeq
That is, the BLM procedure consists of defining the prediction for
${\cal O}$ at this order to be
\beq
{\cal O} = \a_s(\mu^*)\left[1 + \left(A_1+ {33B_1\over2} \right)
{\a_s(\mu^*)\over\pi}  + \cdots\right]\; ,
\eeq
where
\beq
\mu^* \equiv \mu e^{3B_1}\; .
\eeq
Note that knowledge of the NLO term in the expansion is necessary to
fix the scale at LO.  The scale occurring in the highest term in the
expansion will in general be unknown.  A natural prescription is to
set this scale to be the same as that in the next-to-highest-order
term.

A very important feature of this prescription is that $\mu^*$ is
actually independent of $\mu$.  (This follows from considering the
$\mu$ dependence of $B_1$.)  Thus pQCD predictions using the BLM
procedure are unambiguous.

The same basic idea can be extended to higher orders, by
systematically shifting $n_f$ dependence into the renormalization
scales order by order.  The result is that a generic perturbative
expansion
\beq
{\a_s(\mu)\over\pi} 
+ \left(A_1+B_1 n_f\right) \left({\a_s(\mu)\over\pi}\right)^2
+ \left(A_2+B_2 n_f + C_2 n_f^2\right) 
\left({\a_s(\mu)\over\pi}\right)^3 + \cdots
\eeq
is replaced by a series of the form
\beq
{\a_s(\mu^*)\over\pi} 
+ \widetilde{A}_1\left({\a_s(\mu^{**})\over\pi}\right)^2
+ \widetilde{A}_2\left({\a_s(\mu^{***})\over\pi}\right)^3
+ \cdots\; .
\eeq
In general a different scale appears at each order in perturbation
theory.  In addition, the coefficients $\widetilde{A}_n$ are
constructed to be independent of $n_f$, and so the form of the
expansion is unchanged as momenta vary across quark mass thresholds.
All effects due to quark loops in vacuum polarization diagrams are
automatically incorporated into the effective couplings.

As discussed above, one motivation for this prescription is that it
reduces to the correct result in the case of QED.  In addition, when
combined with the idea of commensurate scale relations (see below),
the BLM method can be shown to be consistent with the generalized
renormalization group invariance of St\"uckelberg and Peterman, in
which one considers ``flow equations'' both in $\mu$ and in the
parameters that define the scheme.

To avoid scheme dependence, it is convenient to introduce physical
effective charges, defined via some convenient observable, for use as
an expansion parameter.  An expansion of a physical quantity in terms
of such a charge is a relation between observables and therefore must
be independent of theoretical conventions, such as the renormalization
scheme, to any fixed order of perturbation theory.  In practice, a CSR
relating two observables $A$ and $B$ is obtained by applying BLM
scale-fixing to their respective perturbative predictions in, say, the
$\overline {MS}$ scheme, and then algebraically eliminating
$\alpha_{\overline {MS}}$.

A particularly useful scheme is furnished by the heavy quark potential
$V(Q^2)$, which can be identified as the two-particle-irreducible
amplitude for the scattering of an infinitely heavy quark and
antiquark at momentum transfer $t = -Q^2$.  The relation
\begin{equation}
V(Q^2) = - {4 \pi C_F \alpha_V(Q)\over Q^2}\; ,
\end{equation}
with $C_F=(N_C^2-1)/2 N_C=4/3$, then defines the effective charge
$\alpha_V(Q)$.  This coupling provides a physically-based alternative
to the usual ${\overline {MS}}$ scheme.  Another useful charge is
provided by the total $e^+e^-\to X$ cross section, via the definition
$R(s) \equiv 3\Sigma e_q^2 \left(1 + \alpha_R(\sqrt s)/ \pi\right).$
The CSR relating $\alpha_V$ and $\alpha_R$ is
\begin{equation}
\alpha_{V}(Q_V) = \alpha_R(Q_R) \left(1 - {25\over 12}{
\alpha_R\over\pi} + \cdots\right),
\label{NLOalpv}
\end {equation}
where the ratio of commensurate scales is $Q_R/Q_V =
e^{23/12-{2\zeta_3}} \simeq 0.614$ \cite{bjpr97}.

Physical couplings like $\a_V$ are of course
renormalization-group-invariant, i.e. $\mu\del \a_V/\del\mu = 0$.
However, the dependence of $\a_V(Q)$ on $Q$ is controlled by an
equation which is formally identical to the usual RG equation.  Since
$\a_V$ is dimensionless we must have
\beq
\a_V = \a_V\left({Q\over\mu}, \a_s(\mu)\right)\; .
\eeq
Then $\mu\del\a_V/\del\mu = 0$ implies
\beq
Q{\del\over\del Q}\a_V(Q) = \b_s(\a_s){\del\a_V\over\del\a_s}
\equiv \b_V(\a_V)\; ,	
\eeq
where
\beq
\b_s = \mu{\del\over\del\mu}\a_s(\mu)\; .
\eeq
This is formally a change of scheme, so that the first two
coefficients $\beta_0 = 11 - 2 n_f/3$ and $\beta_1 = 102 - 38n_f/3$ in
the perturbative expansion of $\b_V$ are the standard ones.

\section*{EXCLUSIVE AMPLITUDES AND $V(Q^2)$}

Exclusive processes are particularly challenging to compute in QCD
because of their sensitivity to the unknown nonperturbative bound
state dynamics of the hadrons.  However, there is an extraordinary
simplification which occurs when the hadrons are forced to absorb a
large momentum transfer $Q$: one can separate the nonperturbative
long-distance physics associated with hadron structure from the
short-distance quark-gluon hard scattering amplitudes responsible for
the dynamical reaction.  A meson form factor, for example, factorizes
to leading order in $1/Q$ in the form
\begin{equation}
F_M(Q^2)=\int^1_0 dx \int^1_0 dy \phi_M(x, {\tilde Q}) T_H(x,y,Q^2)
\phi_M(y,{\tilde Q})\; ,
\end{equation}
where $\phi_M(x,{\tilde Q})$ is the meson distribution amplitude,
which encodes the nonperturbative dynamics of the bound valence Fock
state up to the resolution scale ${\tilde Q}$, and $T_H$ is the
leading-twist perturbatively-calculable amplitude for the subprocess
$\gamma^* q(x) \overline q(1-x) \to q(y) \overline q(1-y)$, in which
the incident and final mesons are replaced by valence quarks collinear
up to the resolution scale ${\tilde Q}$.  Contributions from
nonvalence Fock states and the correction from neglecting the
transverse momenta in the subprocess amplitude from the
nonperturbative regime are higher twist, i.e., power-law suppressed.
The transverse momenta in the perturbative domain lead to the
evolution of the distribution amplitude in ${\tilde Q}$ and to NLO
corrections in $\alpha_s$.  For further details and references see
\cite{BrodskyLepage}.

It is straightforward to obtain CSRs relating exclusive amplitudes to,
e.g., the heavy quark potential.  For the pion form factor, for
example, we find \cite{bjpr97}
\begin{equation}
F_\pi(Q^2) =\int^1_0 dx \phi_\pi(x) \int^1_0 dy \phi_\pi(y) 
{4 \pi C_F\alpha_V(Q^*_V)\over (1-x) (1-y) Q^2} \left(1 + C_V
{\alpha_V(Q^*_V))\over\pi}\right)\; ,
\label{pionformfactor}
\end{equation}
where $C_V = -1.91$ and ${Q_V^*}^2 = (1-x)(1-y)Q^2$ is the virtuality
of the exchanged gluon in the underlying hard scattering amplitude.
Eq. (\ref{pionformfactor}) represents a general connection between the
form factor of a bound-state system and the irreducible kernel that
describes the scattering of its constituents.

If we expand the QCD coupling about a fixed point \cite{bjpr97}, and
assume that the pion distribution amplitude has the asymptotic form
$\phi_\pi(x) = {\sqrt 3} f_\pi x (1-x), $ with $f_\pi \simeq 93$ MeV,
then the integral over the effective charge in
Eq. (\ref{pionformfactor}) can be performed explicitly.  We thus find
\begin{equation}
Q^2 F_\pi(Q^2) = 16 \pi f^2 _\pi \alpha_V(e^{-3/2} Q) \left(1 - 1.91
{\alpha_V\over \pi}\right)
\label{pionformfactornlo}
\end {equation}
for the asymptotic distribution amplitude.  In addition
\begin{equation}
Q^2 F_{\gamma \pi}(Q^2)= 2 f_\pi \left( 1 - {5\over3}
{\alpha_V(e^{-3/2}Q)\over \pi}\right)
\label{qsquaretransition}
\end{equation}
for the $\g\to\pi^0$ transition form factor.  A further prediction
resulting from the factorized form of these results is that the
normalization of the ratio
\begin{eqnarray}
R_\pi(Q^2) &\equiv & \frac{F_\pi (Q^2)}{4 \pi Q^2 |F_{\pi
\gamma}(Q^2)|^2}
\label{Rpidef}\\
&=& \alpha_V(e^{-3/2}Q)\left(1+1.43 {\alpha_V\over\pi} \right)
\label{RpiV}
\end{eqnarray}
is formally independent of the form of the pion distribution
amplitude.  The NLO correction given here assumes the asymptotic
distribution amplitude.

A striking feature of these results is that the physical scale
controlling the form factor in the $\alpha_V$ scheme is very low:
$e^{-3/2} Q \simeq 0.22 Q$, reflecting the characteristic momentum
transfer experienced by the spectator valence quark in lepton-meson
elastic scattering.  In order to compare these expressions to data,
therefore, we require an Ansatz for $\a_V$ at low scales.  In
Ref. \cite{bjpr97} we consider a parameterization of the form
\begin{equation}
\alpha_V(Q) = {4 \pi \over {\beta_0 \ln \left({{Q^2 + 4m_g^2}\over
\Lambda^2_V}\right)}} ,
\label{frozencoupling}
\end{equation}
which effectively freezes $\a_V$ to a constant value for $Q^2 \leq
4m_g^2$.  A primary motivation for this is the observation that the
data for exclusive amplitudes such as form factors, two-photon
processes such as $\gamma \gamma \to \pi^+ \pi^-,$ and photoproduction
at fixed $\theta_{c.m.}$ are consistent with the nominal scaling of
the leading-twist QCD predictions for momentum transfers down to a few
GeV.  This can be immediately understood if $\alpha_V$ is slowly
varying at low momentum.  The scaling of the exclusive amplitude then
follows that of the subprocess amplitude $T_H$ with effectively fixed
coupling.

The parameters $\L_V$ and $m_g^2$ are determined by fitting to a
lattice determination of $V(Q^2)$ \cite{Davies} and to a value of
$\a_R$ advocated in \cite{MattinglyStevenson} using
Eq. (\ref{NLOalpv}).  We find $\Lambda_V \simeq 0.16$ GeV and
$m^2_g\simeq 0.2$ GeV$^2$.  With these values, the prediction for
$F_{\g\pi}$ is in excellent agreement with the data for $Q^2$ in the
range 2--10 GeV$^2$.  We also reproduce the scaling and normalization
of the $\gamma\gamma \rightarrow \pi^+\pi^-$ cross section at large
momentum transfer.  However, the normalization of the space-like pion
form factor $F_\pi(Q^2)$ obtained from electroproduction experiments
is somewhat higher than that predicted by
Eq. (\ref{pionformfactornlo}).  This discrepancy may actually be due
to systematic errors introduced by the extrapolation of the $\gamma^*
p \to \pi^+ n$ data to the pion pole.  What is at best measured in
electroproduction is the transition amplitude between a mesonic state
with an effective space-like mass $m^2=t < 0$ and the physical
pion. It is theoretically possible that the off-shell form factor
$F_\pi(Q^2,t)$ is significantly larger than the physical form factor
because of its bias towards more point-like $q \overline q$ valence
configurations in its Fock state structure.  These and related issues
are discussed elsewhere \cite{BR}.

In any case, we find no compelling argument for significant
higher-twist contributions in the few-GeV regime from the hard
scattering amplitude or the endpoint regions, since such corrections
would violate the observed scaling behavior of the data.

\section*{SUMMARY}

As we have emphasized, the $\a_V$ scheme is quite natural when
analyzing QCD processes perturbatively.  By definition, it
automatically incorporates quark (as well as the corresponding gluon)
vacuum polarization contributions into the coupling; thus the
coefficients in a perturbative expansion do not change as momenta vary
across quark mass thresholds.  It is directly connected to one of the
most useful observables in QCD, the heavy quark potential, which is
accessible on the lattice as well as phenomenologically via the
spectrum of heavy quarkonium.  Finally, the scale-setting problem in
QCD appears much less mysterious from this point of view: the scale
appropriate for each appearance of $\alpha_V$ in a Feynman diagram is
just the momentum transfer of the corresponding exchanged
gluon.\footnote{There are complications which arise when gluon
self-couplings are present.  These are discussed in \cite{luthesis}.}
This prescription is equivalent to the BLM procedure.

It may also prove useful in the context of nonperturbative
calculations based on the light-cone formalism.  A light-cone
Hamiltonian expressed in terms of $\a_V$, so that it reproduces
covariant perturbation theory with $\a_V$ appearing inside the
momentum integrals, should be very well suited to studying, e.g.,
heavy quark systems: the effects of the light quarks and higher Fock
state gluons that renormalize the coupling are already contained in
$\alpha_V$.  At high momentum scales $\a_V$ should be computable via
perturbation theory, while at low scales a semi-phenomenological
Ansatz would be necessary.  As we have discussed, exclusive processes
can provide a valuable window for determining $\a_V$ in the low-energy
domain.

It is a pleasure to thank S. J. Brodsky, C.-R. Ji and A. Pang for the
very enjoyable collaboration which led to this work.  I also thank the
organizers, in particular P. Grang\'e, for putting together such a
stimulating and enjoyable meeting.  This work was supported in part by
a grant from the U.S. Department of Energy.

\end{document}